# A Connectionless Interaction Model for Network Spatial Proximity Services

Artem Makarov, Dmitry Namiot

***Abstract*—In this paper, we present a connectionless interaction model for network spatial proximity services. In the network spatial proximity model, the reception of an identification or discovery frame of a nearby wireless node confirms proximity and delivers content at the same time. Current Bluetooth and Wi-Fi specifications and mobile platforms extend this one-way announcement with responses, addressed messages, scheduled group exchange, and relaying, which still do not require a connection. The proposed model describes such interaction independently of a concrete wireless technology. It is based on the roles of nearby nodes, the proximity message as the unit of interaction, twelve pattern properties, and six interaction patterns: announcement, mutual announcement, query-response, addressed exchange, scheduled group exchange, and relay. An exit rule defines when a service should move to a connection-based channel. We analyze how the patterns can be implemented on Android and iOS smartphones through native mechanisms, substitution, composition, embedded bridges, or a step-up to a connection, and we propose a pattern selection method for developers. The model generalizes the earlier network spatial proximity model and refines the interaction primitives of the proximity core. It is illustrated by existing services and systems and supported by published measurements and first-order estimates.**

## I. Introduction

Many mobile services need to know that a user is near a place, an object, or another user, and do not need geographic coordinates for this. In the network spatial proximity model, such a condition is expressed directly through the visibility of wireless nodes [1], [2]. An SSID (Service Set Identifier), a Bluetooth Low Energy (BLE) advertising packet, or a Wi-Fi service record is visible before any connection is established, and its reception both confirms proximity and delivers a short message. Such interaction needs no pairing, no network infrastructure, and no Internet access, and a device does not have to accept connections from unknown devices. Services built on this model, such as presentation broadcasting [3], contact exchange [4], and context-aware browsing [5], use it as a one-way announcement: one device announces, and the nearby devices observe.

Current wireless specifications provide more than one-way announcements without connections. BLE 5.0 added extended and periodic advertising, and BLE 5.4 added Periodic Advertising with Responses (PAwR) and Encrypted Advertising Data (EAD) [6]. Wi-Fi Aware supports follow-up messages between discovered peers, and its version 4.0 adds unsynchronized service discovery (USD) [7]. Electronic shelf labels [8], Matter commissioning [9], and a classroom response system [10] already use these mechanisms to exchange data in both directions without a connection after a one-time setup. Mobile platforms expose parts of them: Android made USD-based discovery public in Android 16 [11], and iOS 26 added a Wi-Fi Aware framework [12].

At the service level, however, we found no model that describes this kind of interaction. The proximity core [13] defines primitives for proximity services and a driver-based, capability-aware architecture. Its mapping to BLE and Wi-Fi technologies showed that the capability boundary is sharper for the interaction primitives than for the discovery primitives. The core does not say which forms of connectionless interaction exist and how each of them can be implemented on phones. Work on association-less communication compares Wi-Fi and BLE mechanisms on embedded devices [14], [15], but it does not consider mobile platforms and their application interfaces.

In this paper, we propose a connectionless interaction model for network spatial proximity services. The paper makes four contributions. First, it defines a model of connectionless interaction with roles, the proximity message, twelve pattern properties, six interaction patterns, and an exit rule to connection-based channels. Second, it relates the model to earlier work: the network spatial proximity model becomes a special case, and the interaction primitives of the proximity core are refined. Third, it analyzes implementation strategies for each pattern on Android and iOS, with first-order estimates and published measurements. Fourth, it gives a pattern selection method for developers and describes twelve existing services and systems in terms of the model.

The rest of the paper is organized as follows. Section II gives the background and related work. Section III presents the model, and Section IV analyzes its implementation on mobile platforms. Section V describes existing services and systems in terms of the model, and Section VI gives the pattern selection method. Section VII discusses the results, and Section VIII concludes the paper.

## II. Background and Related Work

### A. Network spatial proximity

The network spatial proximity model replaces geographic coordinates with the visibility of wireless nodes [1]. A service is linked to wireless nodes, and a mobile device that Artem Makarov – Lomonosov Moscow State University (email: artmakar01@mail.ru).

Dmitry Namiot – Lomonosov Moscow State University (email: dnamiot@gmail.com).

sees a node is near the place, object, or person that the node represents. The nodes can be existing ones, such as Wi-Fi access points, or nodes created for the service, such as BLE beacons or phones that advertise. In the model, data are carried in the identification of a node: an SSID, a BLE advertising packet, or a service record contains a short message, so that the reception of the frame is at the same time a proximity statement and a content delivery. Discovery is separated from any further interaction, which may use a connection or a network service. The model was applied to programming of proximity services [16], to proximity between mobile devices [2], and to a hyper-local Internet [17].

The model was used in a presentation service that announces links to slides in BLE advertising packets [3] and in a classroom response system based on BLE 5.4 PAwR [10]. The proximity core [13] generalizes such services by seven primitives: `fingerprint()`, `broadcast()`, `subscribe()`, `presence()`, `sendTo()/receiveFrom()`, `openChannel()`, and `measureDistance()`. A driver translates the primitives into the mechanisms of a concrete platform, and a capability resolver selects a driver for a service request. The present paper refines the interaction primitives of this core.

### B. Connectionless mechanisms beyond announcement

BLE legacy advertising carries up to 31 bytes in an advertising packet. BLE 5.0 added extended advertising, with up to 1650 bytes of chained data, and periodic advertising, in which a receiver synchronizes to the schedule of an advertiser [6], [18]. BLE 5.4 added PAwR: in each periodic interval the advertiser sends up to 128 subevents, and the synchronized devices reply in up to 255 response slots of a subevent, each with up to 251 bytes. BLE 5.4 also added EAD, which encrypts advertising data with a key shared in advance [6]. BLE 6.0 added decision-based advertising filtering and the monitoring of advertisers, which move parts of receiver-side selection and presence into the controller [19].

Wi-Fi has similar mechanisms. Beacon stuffing places application data into the beacons and probe frames of access points [20]. Wi-Fi Direct service discovery exchanges DNS Service Discovery (DNS-SD) or Universal Plug and Play (UPnP) service records before a group is formed [21]. Wi-Fi Aware, also known as Neighbor Awareness Networking (NAN), provides publish and subscribe discovery with service-specific information and follow-up messages that do not require a data path [22]. Its version 4.0 adds USD, which works without synchronization to the NAN cluster [7]. Bluetooth Mesh relays messages between embedded nodes over an advertising bearer [23], and the dissertation [1] describes relaying on phones through node identification. Together, these mechanisms support responses, addressing, schedules, and relaying, while the data still travel in identification and discovery frames.

### C. Related work

Association-less communication is the closest line of work. Issa et al. compare Wi-Fi beacon stuffing and the BLE connectionless mode for ephemeral interactions in smart environments [14] and survey both technologies [15]. Their measurements are made on embedded devices, and the survey does not analyze the application interfaces of Android and iOS. Our model differs in three points: it is built for smartphones and their public interfaces, it describes interaction patterns and not only data dissemination, and it keeps the proximity meaning of a reception.

Several works use BLE advertising as a data channel between phones. Tsai and Chen propose undirected information dissemination over legacy advertising between background Android applications [24], and BLEWhisperer shows that extended advertising can carry data at up to 236 bytes per second from Android phones [25]. Reliability of advertising-based communication was studied on embedded devices [26], [27]. These works study one mechanism and one pattern, mainly the announcement.

Frameworks for interaction between nearby devices, such as Google Nearby Connections [28] and Apple Multipeer Connectivity [29], hide the wireless mechanisms but establish connections between devices. The Omni framework separates lightweight context, which is carried over discovery channels, from heavyweight data, which uses connections [30]. This separation is close to the exit rule of the proposed model, but Omni does not define connectionless interaction patterns or their implementation strategies.

## III. The Connectionless Interaction Model

The proposed model describes interaction between nearby mobile devices that is carried by identification and discovery frames, without establishing a connection. It does not depend on a concrete wireless technology. The model consists of four parts: the elements (roles and the proximity message), the pattern properties, six interaction patterns, and an exit rule that defines the boundary of connectionless interaction. Fig. 1 shows how these parts relate to each other and to the network spatial proximity model.

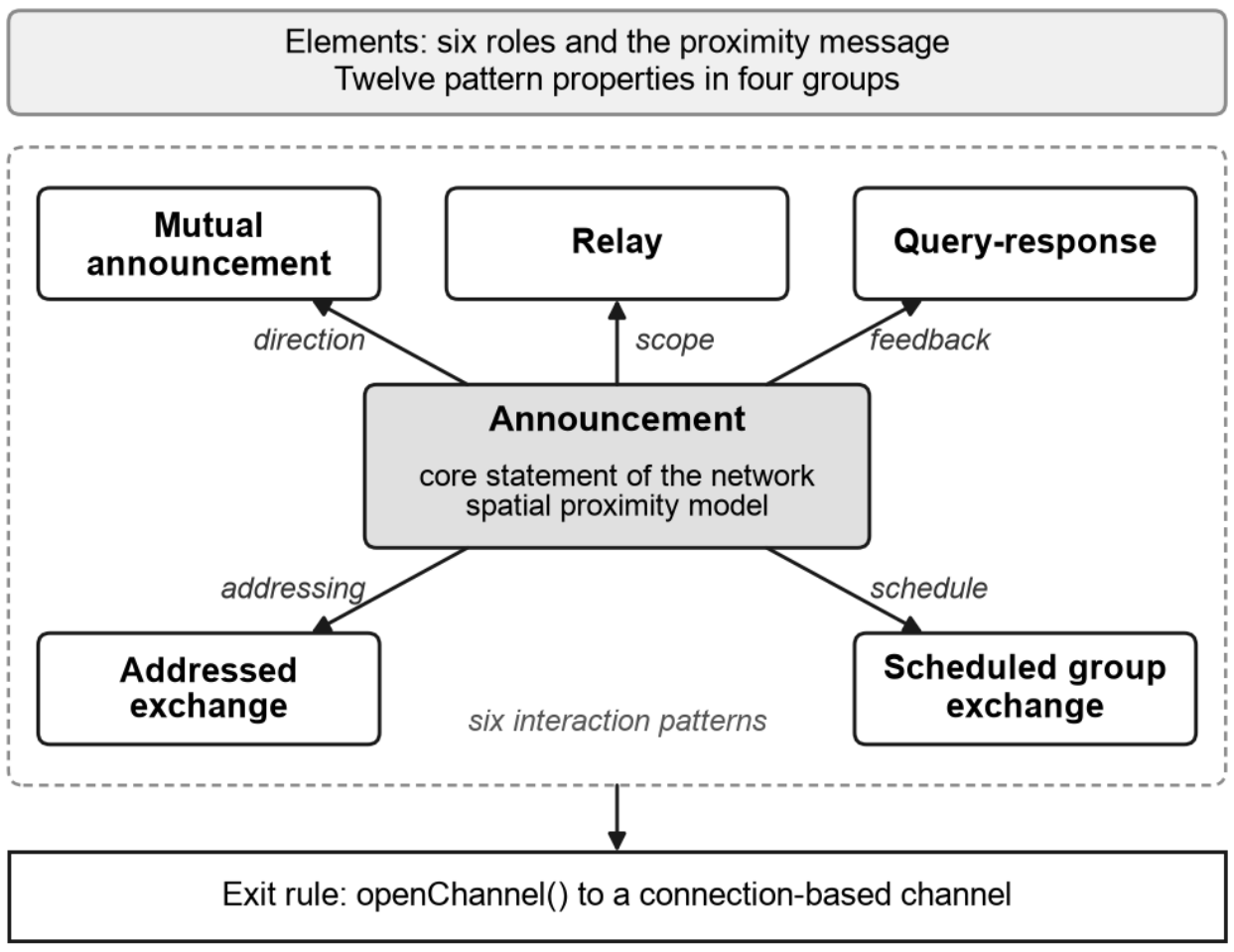

Fig. 1. Overview of the connectionless interaction model. Each of the other patterns changes one property of the announcement pattern.

### A. Elements of the model

The unit of interaction is the *proximity message*. A proximity message is application content placed into an identification or discovery frame of a wireless node, for example into a BLE advertising packet, a Wi-Fi Aware service discovery frame, a Wi-Fi Direct service record, or an SSID. As in the network spatial proximity model [1], the reception of a proximity message has two meanings at once. It states, that the receiver is within radio range of the sender, and it delivers the content. The first meaning is the reason why the message is not just a short datagram: the service logic can use the reception itself as a proximity condition, without any coordinates.

TABLE I
PATTERN PROPERTIES

| Property | Announcement | Mutual announcement | Query-response | Addressed exchange | Scheduled group exchange | Relay |
|---|---|---|---|---|---|---|
| Cardinality and direction | 1:N | N:N | 1:N query, N:1 replies | 1:1 | 1:N, N:1 in slots | 1:N over hops |
| Addressing | none | none | service name (group) | peer identifier | group and participant identifier | none or group |
| Feedback | none | none | response | optional acknowledgment | reply in a slot | none |
| Schedule | random interval | random interval | on request | on request | synchronized subevents | random interval |
| Payload class | identifier to small content | identifier or short message | short message | short message | short message | short message |
| Confidentiality | open | open | open | open or restricted | open or restricted (EAD) | open or restricted |
| Receiver-side selection | application or controller filter | application filter | service name | peer identifier | group and slot | duplicate check, TTL |
| Identifier stability | fixed or rotating | rotating | fixed service name | session identifier | assigned | message identifier |
| Proximity scope | direct | direct | direct | direct | direct | relayed (last hop) |
| Proximity qualification | visibility or signal level | signal level and duration | visibility | visibility or distance | visibility | visibility (last hop) |
| Presence policy | optional | enter and leave | none | none | membership | none |
| Execution conditions | background limits | background scan duty cycle | Wi-Fi state, permissions | pairing on some platforms | coordinator hardware | relays must run |

The model uses six roles. The *announcer* places a proximity message into its frames and repeats it according to a schedule. The *observer* receives proximity messages and selects the relevant ones. The *responder* answers a query inside discovery frames. The *coordinator* defines a common schedule for a group of nodes and addresses its participants. The *relay node* receives a message and announces it again, so that the message reaches nodes outside the range of the original announcer. The *bridge* is usually an embedded device that runs a pattern on behalf of mobile devices that cannot run it through their public interfaces, for example a Bluetooth Mesh proxy node that passes messages of the advertising bearer to a phone over a connection. One device can hold several roles at the same time. In a mutual announcement, every phone is both an announcer and an observer; in a scheduled group exchange, a participant is an observer that replies in its slot, that is, a responder.

The operations of the roles are named in the vocabulary of the proximity core [13]. The existing primitives `broadcast(topic, payload)`, `subscribe(filter)`, `sendTo(peer, bytes)`, `receiveFrom(peer, bytes)`, and `openChannel(peer, opts)` keep their meaning. The model adds the operations that the connectionless patterns need: `query()` and `respond()`, `poll()` and `reply()`, and `relay()`.

### B. Pattern properties

A pattern is described by twelve properties. They form four groups, and Table I gives their values for each pattern.

The first group describes the structure of interaction. *Cardinality and direction* says who sends to whom: one to many (1:N), many to many (N:N), one to one (1:1), or one to many with replies from many to one (N:1). *Addressing* says whether a message is meant for everybody, for a group (for example, the providers of a named service), or for one peer identifier. *Feedback* says whether the sender gets a response, an acknowledgment, or a reply in a reserved slot. *Schedule* says when messages are sent: at random moments within an interval, periodically, in windows synchronized by a coordinator, or on request.

The second group describes the content. *Payload class* gives the order of magnitude of the application payload per frame: an identifier (up to about 32 bytes), a short message (up to about 255 bytes), or small content (up to about 2 KB) that may need several frames. *Confidentiality* says whether the content is open or restricted to the holders of a key. The mechanisms that provide it, such as EAD, are a separate topic. *Receiver-side selection* says where and by which criterion irrelevant messages are dropped: in the application, in the operating system, or in the controller through filters.

The third group describes identity and proximity. *Identifier stability* says whether the sender identifier is fixed, rotating, assigned by the application, or resolvable with a key. *Proximity scope* is direct when the observer receives the message from the announcer itself, and relayed when the message have passed relay nodes. For a relayed message, the proximity statement holds only for the last hop. *Proximity qualification* says what is required to accept proximity: visibility of the frame, a signal level threshold, or a measured distance. The `measureDistance(mode, target)` primitive of the proximity core belongs here. *Presence policy* says whether repeated receptions are turned into enter and leave events, which is the role of `presence(policy)`.

The fourth group consists of one property, *execution conditions*. It covers the requirements of the platform: foreground or background execution, permissions, pairing, the state of the radio, and regional restrictions. These conditions do not change the logic of a pattern, but they decide whether a pattern can run on a given platform and how often a device can send and receive. Section IV uses

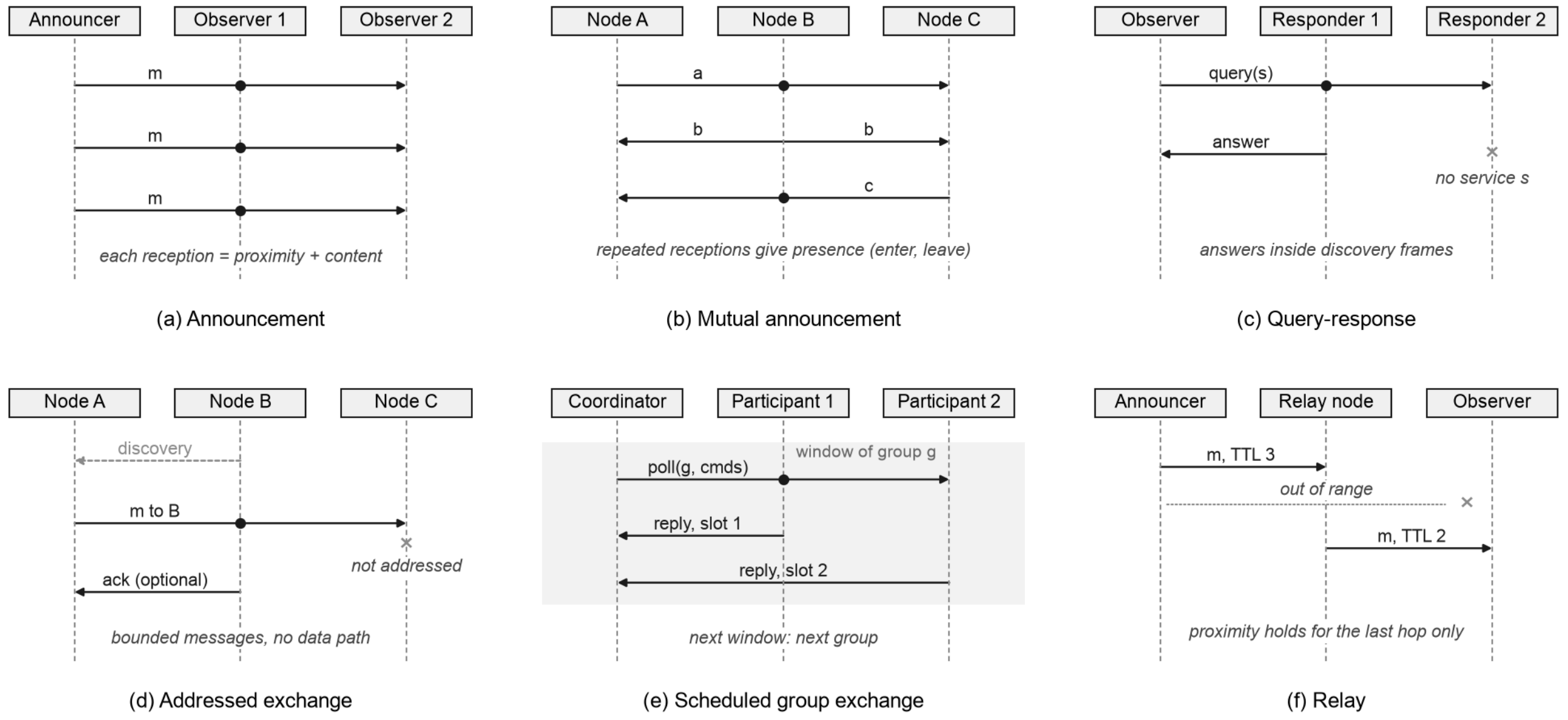

Fig. 2. Message sequences of the six interaction patterns.

them to explain implementation strategies.

## C. *Interaction patterns*

The model defines six patterns. They are grounded in the connectionless mechanisms of current specifications, and the set is open: a new mechanism can add a pattern if it changes the properties in a way that the existing patterns do not cover. Fig. 2 shows the message sequence of each pattern, and the short listings below show how a service uses the operations.

**Announcement.** The announcer encodes its message and repeats it in its frames. Observers filter the frames they receive and decode the relevant messages. Each reception is a proximity statement and a content delivery. BLE calls this advertising and scanning, and Wi-Fi Aware calls it publish and subscribe; the model uses one name for all technologies, and the corresponding operations of the proximity core are `broadcast()` and `subscribe()`. A special case is the observation of nodes that do not take part in the service, such as Wi-Fi access points: their identifiers and signal levels form a fingerprint of the place, which is the `fingerprint()` primitive of the proximity core.

```
announcer: broadcast(topic, payload)
observer:  subscribe(filter) -> (peer, payload)
```

**Mutual announcement.** Every node announces its own message and observes the messages of the others. No node waits for a response, but each node learns about the others in the same way as they learn about it. Repeated receptions from the same sender give presence, and the presence policy turns them into enter and leave events. Contact exchange between nearby users [4] and exposure notification systems (2020–2023) [31] follow this pattern.

```
each node: broadcast(topic, ownPayload)
           subscribe(filter)
           presence(policy) -> enter, leave
```

**Query-response.** The observer announces a query, for example the name of a service. A responder that provides this service answers inside discovery frames, and the observer correlates the answers with its query. The query makes the interaction initiated by the observer, which saves airtime when the responders do not need to announce all the time. Wi-Fi Direct service discovery with DNS-SD records [21] and the unsynchronized service discovery of Wi-Fi Aware [7] follow this pattern.

```
responder: respond(service, q -> answer(q))
observer:  query(service, timeout)
               -> list of (peer, answer)
```

**Addressed exchange.** After discovery, two nodes exchange bounded messages addressed to a peer identifier. The exchange may include an acknowledgment, but it does not create a data path or a session. Follow-up messages of Wi-Fi Aware are the native mechanism for this pattern.

```
node A: sendTo(peerB, bytes) -> ack or timeout
node B: receiveFrom(peerA, bytes)
```

**Scheduled group exchange.** The coordinator announces a schedule and addresses groups of participants in successive time windows. Each participant listens only to the window of its group and replies in the slot that the schedule assigns to it. One coordinator thus collects replies from many participants without a connection to any of them. BLE 5.4 PAwR and the Electronic Shelf Label (ESL) profile follow this pattern; in the ESL specifications, a 7-bit group identifier selects the window and an 8-bit identifier selects the label, which gives up to 32,640 addressable labels [32].

```
coordinator: poll(group, commands) -> replies
participant: on poll(group): reply(slot, bytes)
```

**Relay.** A relay node observes a message, checks whether it has already seen it, decreases its time to live (TTL), and announces it again. The message reaches nodes outside the range of the original announcer, but the proximity statement now holds only for the last hop. The advertising bearer of Bluetooth Mesh follows this pattern on embedded nodes [23], and an earlier phone application re-announced received messages as its own node identification [1].

```
relay node: relay(filter)
  on m: if new(m.id) and m.ttl > 1:
          broadcast(m.topic, m with ttl - 1)
```

## D. *Exit rule*

A connectionless pattern is sufficient while the payload fits its payload class, while losses can be covered by repetition, and while the content does not need a protected session. When one of these conditions fails, the service leaves the pattern through `openChannel(peer, opts)`. The exit starts from the current pattern and uses what it has already established: the peer identifier, the service name, or the schedule. For example, a PAwR coordinator can open a connection to a participant within a subevent [10], a Wi-Fi Aware discovery session can be followed by a data path, and a BLE observer can connect to the announcer it has received. The exit rule therefore marks the boundary of the model, and not a failure of a pattern.

### *E. Relation to earlier models*

The network spatial proximity model [1] is based on one-way data in node identification: the reception of a frame confirms proximity and delivers content, and discovery is separated from any further interaction. In the proposed model this core statement is the announcement pattern with the following property values: one to many, no addressing, no feedback, a random schedule, open content, and a direct proximity scope. Each of the other patterns changes one principal property of this base pattern, and the other changed values in Table I follow from the principal one. The mutual announcement changes the cardinality and direction to many to many, query-response adds feedback, the addressed exchange adds addressing, the scheduled group exchange adds a synchronized schedule (which then also defines group addressing and slot feedback), and the relay changes the proximity scope. The dissertation [1] also describes the relaying of messages through node identification and the query-response discovery of Wi-Fi Direct, but as applications of the model and not as separate patterns with their properties. The proposed model therefore generalizes the earlier one and keeps its main statement: the reception of a proximity message is a proximity statement.

TABLE II
OPERATIONS OF THE PATTERNS AND THEIR RELATION TO THE PROXIMITY CORE

| Operation | Pattern | Relation to the proximity core |
|---|---|---|
| `broadcast()`, `subscribe()` | announcement, mutual announcement | kept |
| `fingerprint()` | announcement | observation of non-participating nodes |
| `presence()` | announcement, mutual announcement | presence policy property |
| `query()`, `respond()` | query-response | new |
| `sendTo()`, `receiveFrom()` | addressed exchange | kept; explicitly connectionless |
| `poll()`, `reply()` | scheduled group exchange | new |
| `relay()` | relay | new |
| `measureDistance()` | any | proximity qualification property |
| `openChannel()` | exit rule | kept |

The proximity core [13] defines seven primitives and a driver-based, capability-aware architecture. The proposed model refines the interaction primitives of the core. The primitives `broadcast()`, `subscribe()`, `sendTo()`, `receiveFrom()`, and `openChannel()` keep their names; `sendTo()` and `receiveFrom()` stay bounded and are now explicitly connectionless. In the core, this pair covered every exchange after discovery, including the PAwR-based service and the local services over Wi-Fi Direct. The model separates three forms of such exchange, which differ in addressing, feedback, and schedule: the addressed exchange keeps `sendTo()` and `receiveFrom()`, query-response gets `query()` and `respond()`, and the scheduled group exchange gets `poll()` and `reply()`. The `relay()` operation is new. The primitives `fingerprint()`, `presence()`, and `measureDistance()` become a special case of the announcement pattern and two pattern properties. Table II lists the operations by pattern and summarizes these changes.

## IV. Implementation on Mobile Platforms

The model does not depend on a wireless technology, but a service runs on concrete devices. This section analyzes how the patterns can be implemented on current smartphones, Android 17 (API 37) and iOS 27, as of November 2026. The analysis does not ask whether a pattern is supported or not. It asks which implementation strategy is available for each pattern and pair of platforms, and what this strategy costs in payload and time.

### *A. Exposure layers*

A connectionless mechanism reaches an application through several layers. The specification defines the feature, the controller of the chip implements it, and the operating system stack uses it. The public application interface then decides whether third-party applications can use the feature, and the execution conditions (foreground or background execution, permissions, pairing) decide when they can use it. The last layer is the reach across platforms: devices of two platforms can interact only through the mechanisms exposed on both of them.

A feature may stop at any of these layers. BLE 5.4 PAwR and EAD [6] are implemented in embedded controllers and used, for example, by electronic shelf labels [8], but neither Android nor iOS exposes them to applications [33], [34]. Android exposes the transmission of periodic advertising but not its reception [33], [35]. USD [7] has only a system interface on Android, while the same discovery is public as a part of Wi-Fi Direct Release 2 (R2) since Android 16 on devices that support R2 [11]. On iOS, the Wi-Fi Aware framework is public since iOS 26, but it requires pairing before any exchange and gives no access to service-specific information or follow-up messages [12]. Fig. 3 shows the delay between the specification of each mechanism and its first public application interface. The transmit and receive sides are often exposed at different times or only on one platform, so the direction of a pattern has to be checked separatley for each side.

### *B. Payload capacity and first-order estimates*

Fig. 4 compares the application payload that one frame can carry. On phones, the values range from about 19 bytes for the local name that an iOS application can announce in the foreground [34] to 1024 bytes of service-specific information in USD-based Wi-Fi Direct discovery [11], [36]. An SSID carries 16 to 32 bytes, depending on whether the receiver accepts arbitrary bytes or only printable text. An Android application can update the data of an enabled extended advertising set with up to 251 bytes [35], an iOS application receives extended advertising packets of up to 124 bytes sent on the LE 2M physical layer (PHY) [37], and Wi-Fi Aware messages are limited to about 255 bytes [38]. These values define the payload class property: identifiers

| | Android-Android | iOS-iOS | Android-iOS | Phone-embedded device |
|---|---|---|---|---|
| **Announcement** | **Native** BLE advertising, Wi-Fi Aware publish, hotspot SSID | **Native** local name and service UUIDs (foreground) | **Native** Android: any advertising data; iOS: local name (foreground) | **Native** BLE advertising |
| **Mutual announcement** | **Native** advertising and scanning | **Native** advertising and scanning (foreground) | **Native** advertising and scanning (foreground on iOS) | **Native** advertising and scanning |
| **Query-response** | **Native** Wi-Fi Direct DNS-SD; USD on R2 devices | **Substitution** Bonjour over Apple peer-to-peer Wi-Fi | **Composition** query and answer as correlated announcements | **Composition** correlated BLE announcements; USD with R2 phones |
| **Addressed exchange** | **Native** Wi-Fi Aware follow-up messages | **Step-up** paired Wi-Fi Aware data path (composition in the foreground) | **Composition** announcement with an addressee tag | **Native** Wi-Fi Aware follow-up messages (Android); composition with iOS |
| **Scheduled group exchange** | **Composition** announced time slots | **Composition** announced time slots (foreground) | **Composition** announced time slots | **Bridge** PAwR on embedded nodes; phone through a coordinator node |
| **Relay** | **Composition** re-announcement with TTL and a duplicate check | **Composition** re-announcement (foreground) | **Composition** re-announcement with TTL | **Bridge** Mesh advertising bearer; phone through a GATT proxy |

Legend: Native, Substitution, Composition, Bridge, Step-up

Fig. 5. Implementation strategies of the patterns on Android 17 and iOS 27.

fit every mechanism, short messages need a carrier of about 255 bytes, and small content needs fragmentation. They also give the first quantitative bounds for the capability profiles of the proximity core, which were qualitative [13].

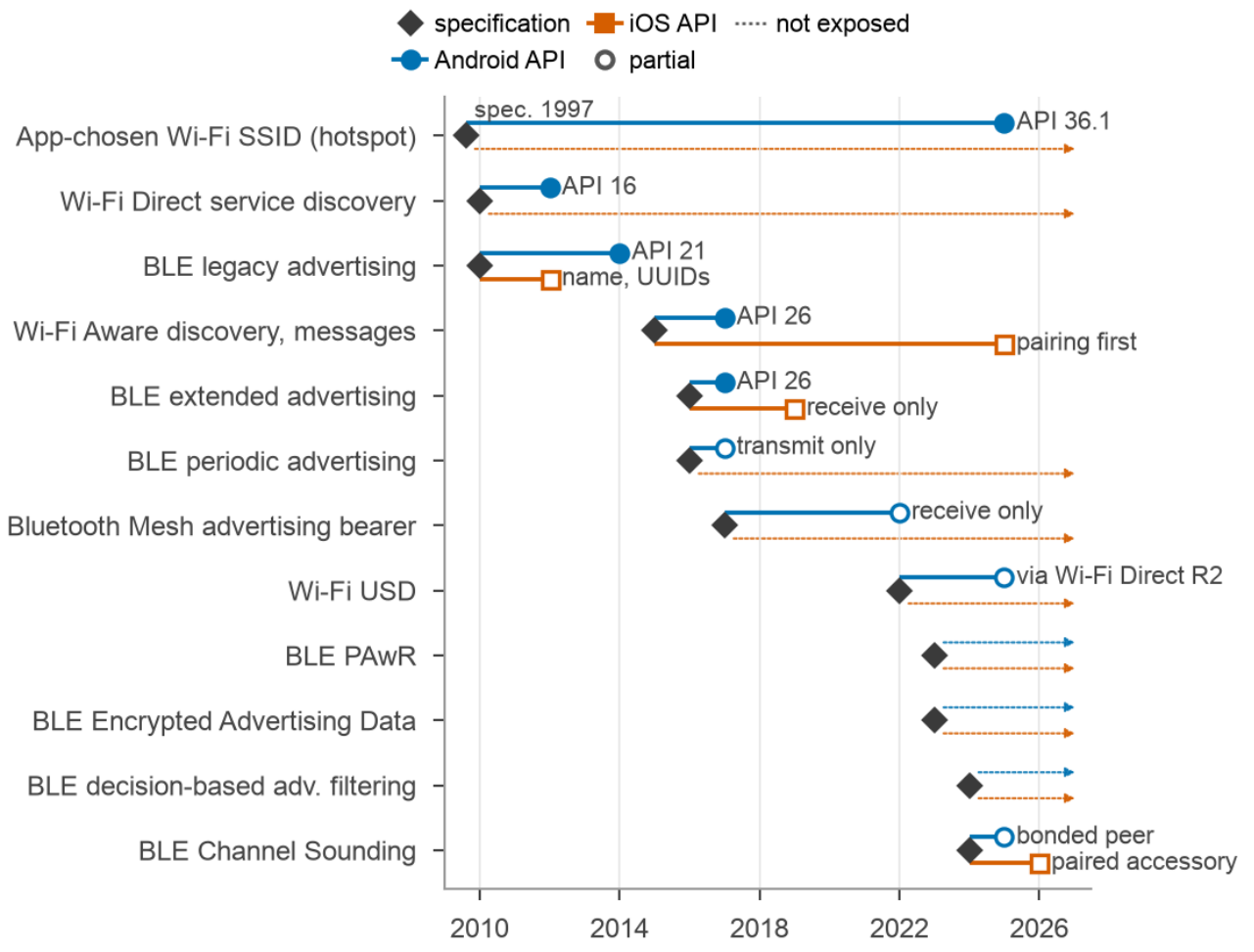

Fig. 3. Year of the specification and of the first public application interface on Android and iOS (November 2026).

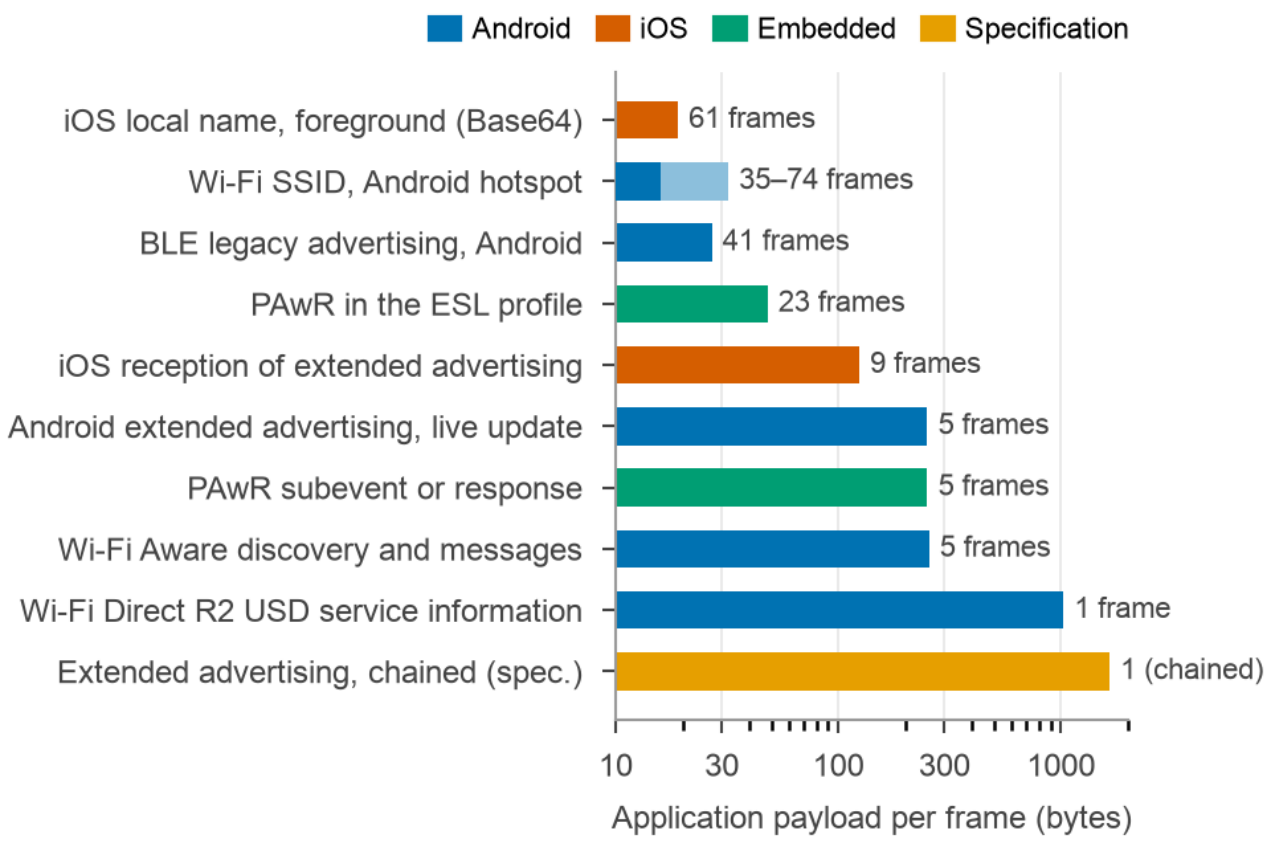

Fig. 4. Application payload per frame of connectionless mechanisms. The labels give the number of frames needed to deliver 1 KB.

For content of size $S$, a payload of $p$ bytes per frame, and a fragment header of $h$ bytes, the announcer needs $n = \lceil S / (p - h) \rceil$ frames, and one pass over all fragments takes $n \cdot d$, where $d$ is the time per fragment. For 1 KB and $h = 2$, $n$ is 41 for legacy advertising ($p = 27$), 61 for the iOS local name ($p = 19$), 9 for iOS reception of extended advertising ($p = 124$), 5 for extended advertising ($p = 251$) and Wi-Fi Aware ($p = 255$), and 1 for USD ($p = 1024$), because content that fits into one frame needs no fragment header. On the transmit side, $d$ can be short: Android does not limit the rate of advertising data updates [33]. On the receive side, a phone scanner receives only a part of the frames, so the effective $d$ is larger. In the presentation service [3], 750 bytes in 50 legacy fragments were delivered between two Android 14 phones at 10 m in about 22.5 s in the best case and 55 s in the worst case, that is, 0.45–1.1 s per fragment against 150 ms scheduled. With extended advertising, the same content in 4 fragments took about 0.4 s and 1.4 s.

In announcement patterns, many announcers share the same advertising channels. The collision model of Ghamari et al. [39] gives the probability that a frame is not lost in a collision as $q = \exp(-2N\lambda d_p)$, where $N$ is the number of announcers, $\lambda$ the rate of advertising events, and $d_p$ the frame duration (376 µs for a 47-byte legacy frame). With $k$ repetitions of the same content, the probability to receive it at least once is $1 - (1 - q)^k$. For 210 announcers with a 250 ms interval and a mean random delay of 5 ms, $q \approx 0.54$, and five repetitions give 0.98. Nikodem and Bawiec measured a collision probability of 0.22–0.33 for this case and a data reception rate above 99% for data intervals longer than 4 s [26], so the estimate is conservative with respect to collisions, while other losses, mainly at the receiver, lower the delivery of single frames further. For announcement patterns, repetition is therefore the main means of reliable delivery, and the schedule property decides its cost.

### C. *Implementation strategies*

The model uses five implementation strategies. In the proximity core, a driver may support a primitive directly, by composition, or only approximately [13]; the strategies make this choice explicit for each pattern and pair of platforms. A pattern is implemented *natively* when the platform exposes a mechanism with the required properties. *Substitution* uses another mechanism of the same platform that gives the same properties, for example Bonjour service records over Apple peer-to-peer Wi-Fi instead of Wi-Fi Aware service discovery. *Composition* builds the pattern from other patterns at the application level, usually from announcements in both directions: addressing is replaced by an addressee tag and receiver-side selection, and a schedule by announced time slots. A *bridge* is an embedded device that runs the pattern natively, while the phone uses it through another pattern or a connection. A *step-up* replaces the pattern by a connection-based mechanism, that is, it applies the exit rule earlier. Fig. 5 gives the strategy for each pattern and pair of platforms. In most cells it is the first available strategy in this order; a later one is chosen when an earlier one cannot keep a property that the pattern requires, as explained below.

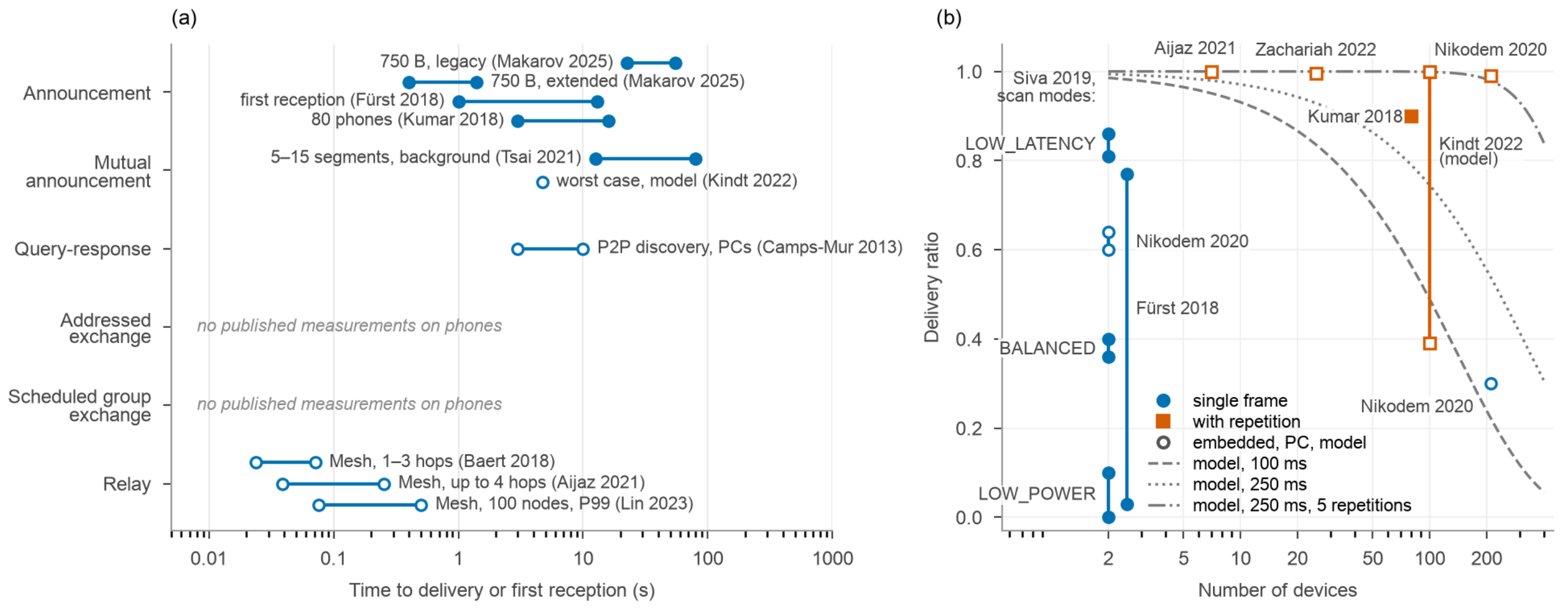

Fig. 6. Indicative values from published measurements: (a) time to delivery or to the first reception; (b) delivery ratio against the number of devices. Filled markers are phones; hollow markers are embedded devices, PCs, or models. Pairwise points in (b) are offset horizontally.

The announcement and mutual announcement are native on both platforms and across them, with different limits. iOS announces only a local name and service identifiers, in the foreground [34], but it receives manufacturer data, service data, and extended advertising packets [37]. Therefore, an iOS application carries its message in the local name, while an Android application can use any advertising data. Query-response is native on Android through Wi-Fi Direct service discovery and, on R2 devices, through USD [11]; between iOS devices it is substituted by Bonjour over Apple peer-to-peer Wi-Fi [40]; across platforms the query and the answer are composed as two correlated announcements.

The addressed exchange is native on Android through Wi-Fi Aware follow-up messages, which report delivery, although delivery is not guaranteed, and do not require a data path [38]; across platforms it is composed from announcements with an addressee tag. Between iOS devices the same composition works only in the foreground, so a service that needs the exchange in the background steps up to a paired Wi-Fi Aware data path [12].

The scheduled group exchange has no public PAwR interface on phones, so it runs natively on embedded devices, as in the classroom response system [10] and electronic shelf labels [8], and reaches a phone through a bridge. Between phones it is composed from announcements in time slots announced by the coordinator; the slots are then as coarse as the advertising interval and the scan duty cycle, which is why the synchronized schedule with embedded devices needs a bridge. The relay is native on embedded Bluetooth Mesh nodes [23] and composed on phones by re-announcement with a TTL and a duplicate check [1]. SafePath uses re-announcement by phones for pedestrian warnings in SSIDs and doubles the line-of-sight range from 60 m to 120 m [41]. A phone joins an existing Bluetooth Mesh network through a Generic Attribute Profile (GATT) proxy, which is a bridge with a step-up.

The execution conditions decide how often a pattern can run rather than whether it can run. An Android application can start at most five scans in 30 seconds by default [33], scanning in the background has a lower duty cycle [24], and an iOS application in the background does not announce its local name [34]. A service therefore declares these conditions as a part of its pattern and chooses the schedule accordingly.

### D. *Published measurements*

Fig. 6 collects indicative values from published measurements on phones and embedded devices [3], [21], [24], [26], [27], [42]–[48]. They come from diferent setups and is not directly comparable, but they show the scale of the values for each pattern. Phone measurements exist for the announcement and mutual announcement, and for the relay only as a range; query-response was measured on PCs, while for the addressed exchange and the scheduled group exchange on phones and for iOS in general we found no published measurements. On phones, the reception of a single frame is low and varies between devices: from about 3% to 77% across ten phone models [42], and 81–86%, 36–40%, and 0–10% for the three Android scan modes [48]. With repetition, the delivery of content can reach 99% and more in controlled deployments [26], [27], while real deployments reported 50–80% [27]. Delivery times on phones are seconds or tens of seconds: about 1–13 s to the first reception [42] and 12.5–79.6 s for messages of 5–15 segments between background applications [24]. Embedded relays work in tens to hundreds of milliseconds [45]–[47]. These values agree with the estimates above.

## V. Validation Through Existing Services and Systems

The model is validated retrospectively, in the same way as the proximity core [13]. Services and systems that were built independently of the model are described in its terms. The test is whether each of them can be expressed by the six patterns and their properties, with one of the implementation strategies of Section IV, and without new elements. Table III lists the cases, which come from three groups: the services of the authors, the services of the network spatial proximity line, and external systems. The four cases that were used for the proximity core [13] are among them.

TABLE III
EXISTING SERVICES AND SYSTEMS DESCRIBED BY THE MODEL

| Case | Patterns | Mechanism | Implementation | Source |
|---|---|---|---|---|
| Presentation broadcast service | announcement | BLE legacy and extended advertising with cyclic fragments | native (Android) | [3] |
| Classroom response system | scheduled group exchange | BLE 5.4 PAwR subevents and response slots | native (embedded boards) | [10] |
| Quack contact exchange | mutual announcement | BLE service data in 31-byte packets | native (Android) | [4] |
| Local services over Wi-Fi Direct | query-response | DNS-SD service records before group formation | native (Android) | [49] |
| Context-aware browsing | announcement (fingerprint) | Wi-Fi scan of non-participating access points | native (Android) | [5] |
| Bluetooth Radio messaging | relay | received messages re-announced as node names, with a duplicate check | composition (Android) | [1] |
| Exposure notifications (2020–2023) | mutual announcement | BLE service data with rotating identifiers | native (operating system service) | [31], [50] |
| Electronic shelf labels | scheduled group exchange | PAwR with group and label identifiers, EAD | native (embedded) | [8] |
| Matter commissioning | query-response | Wi-Fi USD | native (embedded); phone side through the operating system | [9] |
| SafePath pedestrian warnings | announcement, relay | SSIDs re-announced by phones | composition (Android) | [41] |
| BeaconNet classroom network | query-response | SSIDs with acknowledgment bits (one to many), Bluetooth device names (many to one) | composition and substitution (Android) | [44] |
| Bluetooth Mesh | relay | advertising bearer; GATT proxy for phones | native (embedded); bridge for phones | [23], [45] |

The two services of the authors use the patterns at the opposite ends of the model. The presentation broadcast service [3] is the announcement pattern in its basic form: the speaker device announces a short message, and every listener that receives it is near the speaker and gets the content. The fragmentation measurements of this service given in Section IV extend the same pattern from identifiers to small content. The classroom response system [10] is the scheduled group exchange: the speaker device is the coordinator that addresses groups of participants in PAwR subevents, and participants reply in response slots. Because phones do not expose PAwR, the prototype runs natively on embedded boards, and a phone would join it through a bridge. Its temporary connection for synchronization, and the connection from a subevent that it considers for larger content, are examples of the exit rule.

The services of the network spatial proximity line were described in their papers as separate applications of one idea. The model places them into different patterns and makes the differences explicit. Quack exchanges contact cards in BLE service data, and every phone both announces and observes [4], which is the mutual announcement. Local services over Wi-Fi Direct answer service queries before any group is formed [49], which is query-response. Context-aware browsing observes access points that do not take part in the service [5], which is the fingerprint case of the announcement pattern. Bluetooth Radio re-announces received messages as its own node name and checks duplicates by the originator and a sequence number [1], which is the relay pattern composed on phones.

The external systems show that the same patterns are used outside research prototypes. Electronic shelf labels and Matter commissioning use the scheduled group exchange and query-response on embedded devices [8], [9]. SafePath and BeaconNet implement the announcement, the relay, and query-response on unmodified phones through composition and substitution [41], [44]. Both relied on hotspot interfaces that later became unavailable to applications [3]; since Android 16 QPR2, an application can again set the SSID of a local-only hotspot [51]. In Bluetooth Mesh, as in the relay pattern, the proximity statement holds only for the last hop [45]. None of the cases needed a pattern outside the six, and each of them uses one of the five implementation strategies.

## VI. PATTERN SELECTION METHOD

The model gives the developer of a proximity service a short procedure for choosing a pattern and an implementation strategy (Fig. 7). The procedure starts from the requirements of the service and not from a wireless technology.

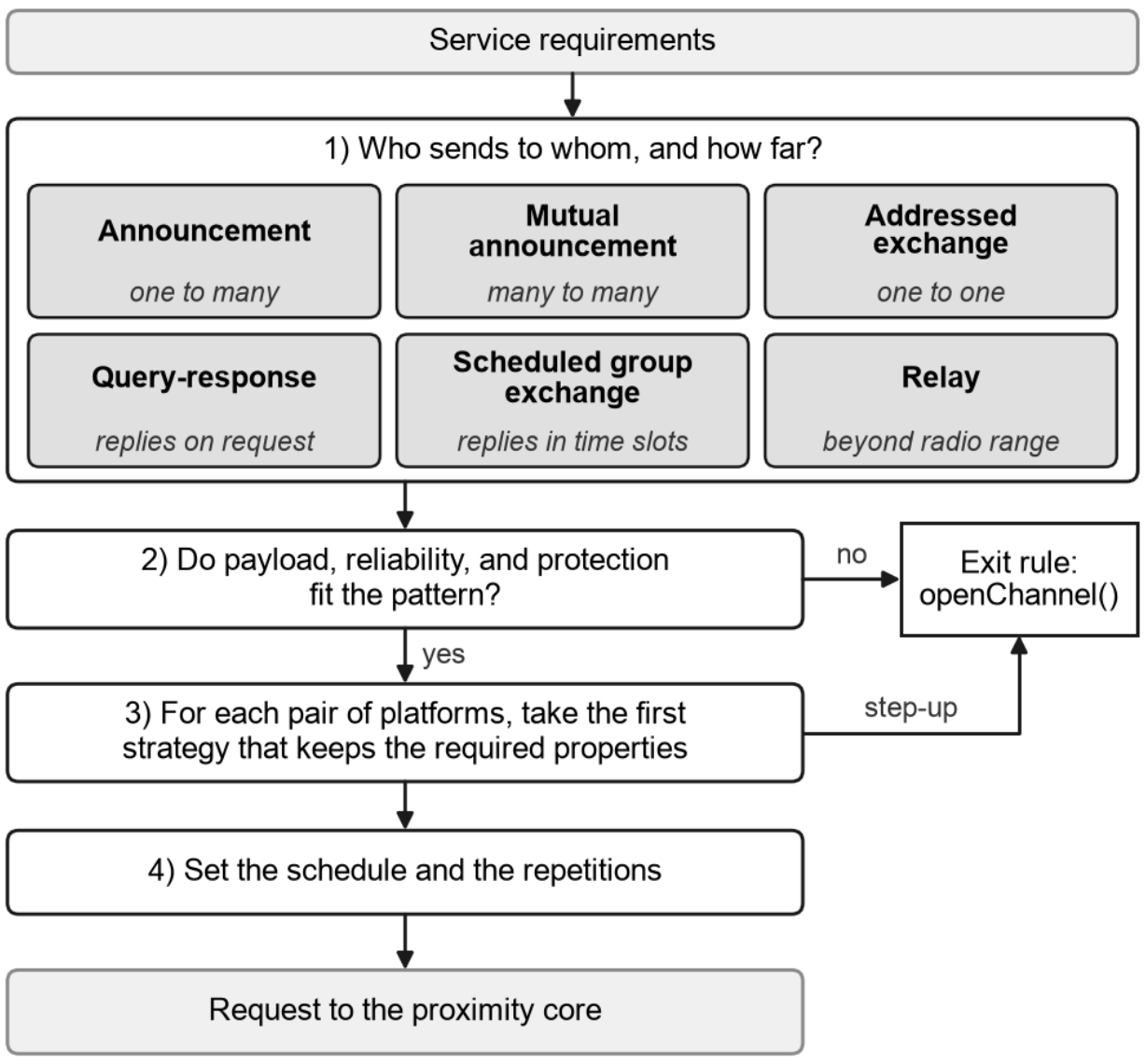

Fig. 7. Pattern selection method.

1) The developer states who sends to whom and how far the message must go. One to many leads to the

announcement, many to many to the mutual announcement, and one to one to the addressed exchange. Replies take precedence over direction: replies on request lead to query-response, replies from many participants within a bounded time to the scheduled group exchange, and delivery beyond radio range leads to the relay.

2) The developer checks whether the payload, the required reliability, and the need for a protected session fit the payload class and the properties of the chosen pattern. If they do not, the service applies the exit rule and opens a channel.

3) For each pair of platforms that must interact, the developer takes the first available strategy that keeps the property values required by the service, in the order native, substitution, composition, bridge, and step-up (Fig. 5). The order follows the growing cost: a substitution changes only the carrier, a composition adds logic to the application, a bridge adds a device, and a step-up adds a connection.

4) The developer sets the schedule and the number of repetitions under the execution conditions, using the estimates of Section IV.

For example, a check-in service for a meeting room that must work between Android and iOS phones is described as follows:

```
service check-in:
  direction = replies on request
  pattern   = query-response, short message
  Android-Android: native (Wi-Fi Direct DNS-SD)
  iOS-iOS: substitution (Bonjour)
  Android-iOS: composition (announcements)
```

In the driver-based architecture of the proximity core [13], the result of this procedure is the request that the capability resolver receives. The service requests a pattern with property values instead of a technology, the resolver selects a driver that implements the pattern with one of the five strategies, a bridge is a driver that works with an embedded device, and the exit rule is the `openChannel()` primitive. The property values express the capabilities that the request requires. The model thus refines the interaction primitives of the core without changing its architecture.

## VII. Discussion

A service is described by a pattern and the values of its properties, the implementation on each pair of platforms is chosen from five strategies, and the exit rule marks where connectionless interaction ends. A gap in a platform interface therefore does not make a service impossible: it changes the strategy and its cost. When a platform exposes a new mechanism, for example PAwR on phones, the corresponding cells of Fig. 5 change from bridge or composition to native, while the description of the service stays the same.

The scope of the paper is the model and its analysis. The platform analysis is based on the documentation and the platform sources of Android 17 and iOS 27 as of November 2026. The interfaces of both platforms change every year; the exposure layers show where such a change belongs, and only the cells of Fig. 5 have to be updated. The services and systems of Section V were built independently of the model, which shows that the model describes existing practice. Confidentiality and identifier stability are properties of the model, and the mechanisms behind them, such as the distribution of keys for encrypted announcements and the resolution of rotating identifiers by authorized observers, are a separate topic.

The next steps follow from this scope. The first is a capability exposure model, in which the layers of Section IV are described for each mechanism and used by the capability resolver of the proximity core. The second is the security of proximity identifiers and messages. The third is the composition of patterns within one service, together with a reference implementation of the corresponding drivers and measurements of the composed and cross-platform strategies on devices.

## VIII. Conclusion

In this paper we presented a connectionless interaction model for network spatial proximity services. The model describes interaction that is carried by identification and discovery frames through roles, the proximity message, twelve pattern properties, six interaction patterns, and an exit rule. The core statement of the network spatial proximity model is the announcement pattern of this model, and each of the other patterns changes one of its principal properties. For current Android and iOS smartphones, every pattern has an implementation strategy for every pair of platforms; between phones, only the addressed exchange between iOS devices in the background needs a step-up to a connection. First-order estimates and published measurements show that timing is set by the schedule and the execution conditions, and delivery by repetition. Twelve existing services and systems were described by the model without new elements, and a selection method connects the model with the capability resolver of the proximity core.